\documentclass{article}

\usepackage{PRIMEarxiv}

\usepackage[utf8]{inputenc} 
\usepackage[T1]{fontenc}    
\usepackage{hyperref}       
\usepackage{url}            
\usepackage{booktabs}       
\usepackage{amsfonts}       
\usepackage{nicefrac}       
\usepackage{microtype}      
\usepackage{lipsum}
\usepackage{fancyhdr}       
\usepackage{graphicx}       
\graphicspath{{media/}}     

\usepackage{color}
\usepackage{algorithm}  
\usepackage{algorithmicx}
\usepackage{algpseudocode}  
\usepackage{amsmath, bm} 
\usepackage{multirow}
\usepackage{makecell}
\usepackage{threeparttable}

\floatname{algorithm}{Algorithm}

\title{SPLDExtraTrees: Robust machine learning approach for predicting kinase inhibitor resistance}

\author{
  Zi-Yi Yang \\
  Tencent Quantum Laboratory\\
  Shenzhen 518057 \\
  China\\
  \texttt{chriszyyang@tencent.com} \\
   \And
  Zhao-Feng Ye \\
  Tencent Quantum Laboratory\\
  Shenzhen 518057 \\
  China\\
  \texttt{michaelzfye@tencent.com} \\
  \And
  Yi-Jia Xiao \\
  Department of Computer Science and Technology\\
  Tsinghua University \\
  Beijing 100084 \\
  China\\
  \texttt{xiaoyiji18@mails.tsinghua.edu.cn} \\
  \And
  Chang-Yu Hsieh \\
  Tencent Quantum Laboratory\\
  Shenzhen 518057 \\
  China\\
  \texttt{kimhsieh@tencent.com} \\
  \And
  Sheng-Yu Zhang \\
  Tencent Quantum Laboratory\\
  Shenzhen 518057 \\
  China\\
  \texttt{shengyzhang@tencent.com} \\
}

\begin{document}
\maketitle

\begin{abstract}
Drug resistance is a major threat to the global health and a significant concern throughout the clinical treatment of diseases and drug development. The mutation in proteins that is related to drug binding is a common cause for adaptive drug resistance. Therefore, quantitative estimations of how mutations would affect the interaction between a drug and the target protein would be of vital significance for the drug development and the clinical practice. Computational methods that rely on molecular dynamics simulations, Rosetta protocols, as well as machine learning methods have been proven to be capable of predicting ligand affinity changes upon protein mutation. However, the severely limited sample size and heavy noise induced overfitting and generalization issues have impeded wide adoption of machine learning for studying drug resistance. In this paper, we propose a robust machine learning method, termed SPLDExtraTrees, which can accurately predict ligand binding affinity changes upon protein mutation and identify resistance-causing mutations. Especially, the proposed method ranks training data following a specific scheme that starts with easy-to-learn samples and gradually incorporates harder and diverse samples into the training, and then iterates between sample weight recalculations and model updates. In addition, we calculate additional physics-based structural features to provide the machine learning model with the valuable domain knowledge on proteins for this data-limited predictive tasks. The experiments substantiate the capability of the proposed method for predicting kinase inhibitor resistance under three scenarios, and achieves predictive accuracy comparable to that of molecular dynamics and Rosetta methods with much less computational costs.

\end{abstract}

\keywords{Drug resistance \and Protein mutation \and Self-paced learning}

\section{Introduction}
In recent decades, drug resistance has been one of the major challenges in the development of anticancer and antimicrobial therapeutics. The resistance can arise via a variety of mechanisms, such as increased drug efflux, drug deactivation, alternative signaling pathways activation and protein mutations. Particularly, protein mutations that directly affect drug binding are found to be the most common mechanism \cite{lovly2014molecular, housman2014drug, ward2020challenges}. On the one hand, the adaptive drug resistance is responsible for the failure of chemotherapy in many advanced cancers. The demands of finding alternative treatment and identifying new drugs that target mutated proteins have been steadily increasing. For example, it is beneficial for drug development by conducting parallel explorations of inhibitors with different resistance profiles \cite{aldeghi2019predicting}. On the other hand, the rapid development of gene sequencing technologies enables the precision medicine to provide personalized recommendations for selecting suitable treatment plans based on the patient's genotype \cite{zehir2017mutational, fowler2018robust}. All these emerging trends urge a better understanding and modeling of the drug resistances and protein mutation, which directly contributes to overcoming the resistances.

Protein kinases play a fundamental role in human cancer initiation and progression \cite{bhullar2018kinase} and are among the most important drug targets \cite{hauser2018predicting}. Many small-molecule kinase inhibitors have been discovered for the treatment of diverse types of cancers by targeting different kinases. A total of more than 62 drugs inhibiting about 20 targets have been approved by the FDA for the treatment of malignancies \cite{roskoski2021properties}. Among them, most inhibitors target tyrosine kinases, which play key roles in the modulation of growth factor signaling and cell proliferation \cite{arora2005role}. Tyrosine kinase inhibitors (TKI) are advantageous in the treatment of chronic myeloid leukemia (CML) and non-small cell lung cancer due to their high efficacy, high selectivity, and low side effects. Some TKIs even have become the first-line drugs for treatment \cite{arora2005role,pottier2020tyrosine}. Despite the successes of TKIs, the emergence of drug resistance remains a major challenge in cancer treatment and has forced the continuous development of new generations of inhibitors \cite{weisberg2007second, y2011recent, juchum2015fighting, neel2017resistance}. Shockingly, over 25\% of CML patients are reported to change TKIs at least once during the treatment due to TKI resistance, most of which are caused by kinases Abl mutations \cite{arora2005role, patel2017mechanisms}. In addition, most of the clinically observed kinase mutations display a long tail of rare and uncharacterized mutations \cite{hauser2018predicting}, which make the sensitivity of known TKIs to these kinase mutants unknown \cite{hauser2018predicting}. The high diversity of kinase mutants makes experimentally examining the influences on TKI time-consuming, expensive and less feasible. Therefore, computational methods are expected to give a clue of how many mutations in target proteins would affect the sensitivity of TKIs. 

Nowadays, three types of computational methods are commonly employed to estimate the affinity changes ($\Delta \Delta$G) of proteins with point mutations \cite{aldeghi2019predicting}. The first is molecular dynamics (MD) based free energy calculations. These calculations use the first-principles statistical mechanics, employing unphysical (i.e., alchemical) intermediates to estimate the free energies changes of the protein thermostability upon a site-mutation \cite{gapsys2015pmx, wang2015accurate, steinbrecher2015accurate, aldeghi2018accurate}. Despite the remarkable performance as a gold standard, it suffers from high computational overhead. The second option is Rosetta, a modeling program that applies mixed physics- and knowledge-based potentials \cite{alford2017rosetta, barlow2018flex} for scoring binding strength. Compared with MD-based simulations, Rosetta with the REF15 scoring function achieves a relative balance between accuracy and computational cost. The third one is the machine learning method \cite{aldeghi2019predicting}, which is data-driven and uses expert-engineered features for predicting $\Delta \Delta$G. For example, Aldeghi et al. employ extremely randomized regression trees to predict TKI affinity change values. Once the input features related to the change in protein mutation affinity are calculated, the change in binding affinity can be predicted within a few seconds. This enables high-throughput evaluations on a wide-range of protein mutations that would potentially induce drug resistance. Meanwhile, this would also be beneficial to the discovery of new drugs targeting mutant proteins.

Although machine learning holds a significant speed advantage over other computational methods, it suffers from the overfitting issue in predicting changes in protein affinity \cite{aldeghi2019predicting}. The main reason is that machine learning usually requires a large amount of labeled training data, which can be impractical or expensive to acquire. Limited training data tends to make predictions become more challenging. Besides, random noise or system and collection biases are inevitably present in the samples, which tends to induce overfitting issue and lead to poor generalization performance. Sample reweighting approach is a commonly used strategy against this robust learning issue \cite{shu2019meta, yang2020smspl}, whose main idea is to impose weights on samples based on their reliability for training. Typical methods include self-paced learning (SPL) \cite{kumar2010self} and variants \cite{jiang2014easy, jiang2014self, yang2019multi, yang2020smspl}.

To alleviate the overfitting issue, we present a robust learning strategy, extremely randomized regression trees with diversity self-paced learning (SPLDExtraTrees), to predict ligand binding affinity changes upon protein mutation for the cancer target Abl kinase. It incorporates a sample reweighting strategy into the extremely randomized regression trees to adaptively suppress the negative influence of noisy data. Specifically, the proposed method selects not only easy samples (with the small loss values), but also focus on balancing sample diversity (from different protein families). Incorporating the diversity of protein families is expected to help quickly grasp easy and comprehensive information and to achieve improved generalization performance. In addition, we also incorporate additional physics-based structural features using Rosetta REF15 scoring function to capture a more sophisticated protein knowledge compared with the work in \cite{aldeghi2019predicting}. We systematically evaluate the capability of SPLDExtraTrees by conducting experiments on human kinase Abl dataset under three scenarios. The experiments demonstrate the potential of SPLDExtraTrees for predicting ligand binding affinity changes upon protein mutations for Abl kinase. It outperforms state-of-the-art machine learning methods by a large margin under all competitive scenarios, and achieves predictive accuracy comparable to that of molecular dynamics and Rosetta methods with less computational time and resources. We expect that the proposed method would effectively provide routine prediction of resistance-causing mutations across other protein targets.

\section{Materials and Methods}
\begin{figure*}[t]
\centering
\includegraphics[width=0.9\textwidth]{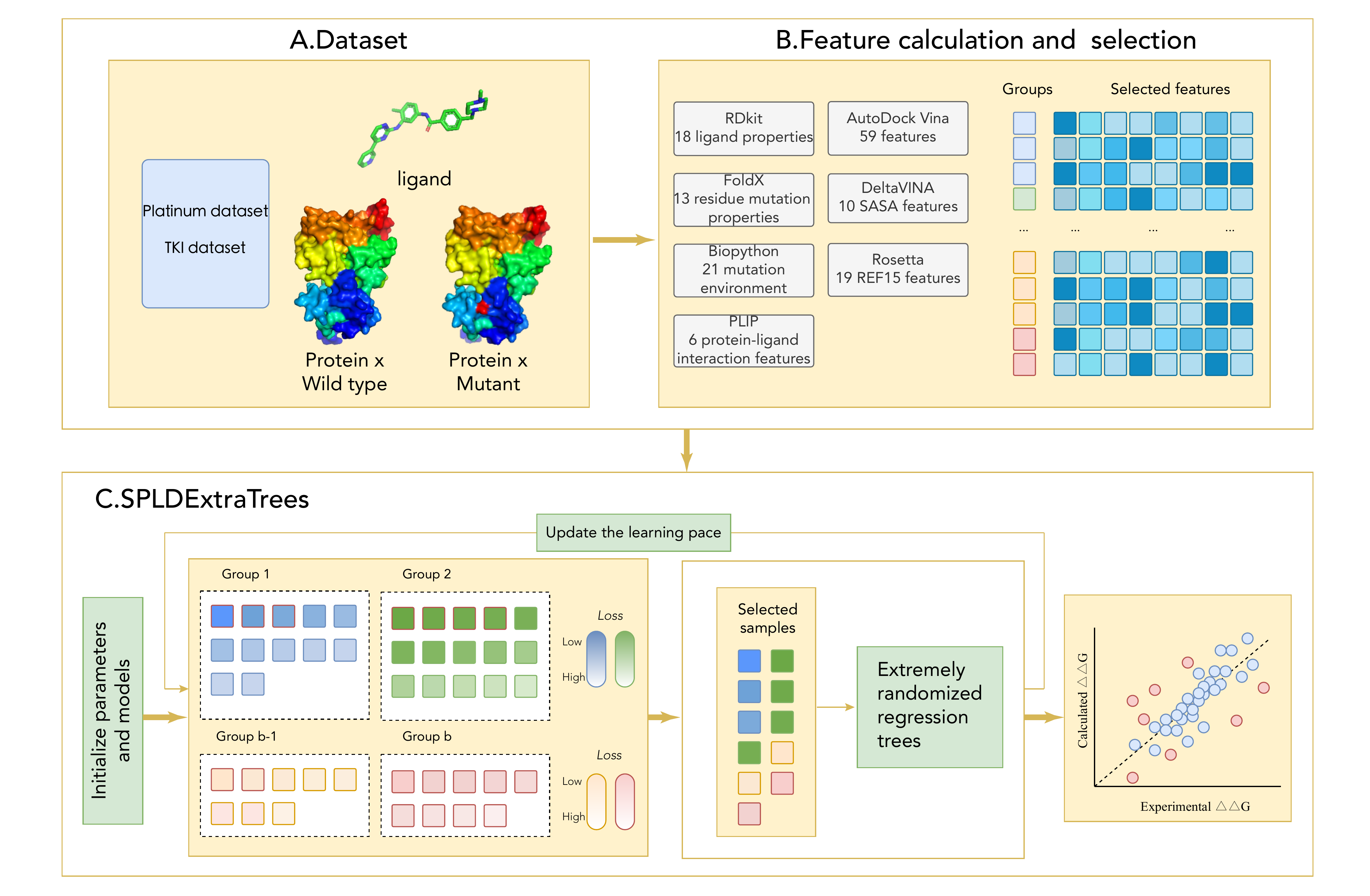}
\caption{Schematic representation of the SPLDExtraTrees workflow. (A) The platinum dataset and TKI dataset are used in this work. Given wild-type, mutant proteins and ligand structure files. (B) Feature calculation and selection. A total of 146 features are calculated, which might contain useful information for predicting changes in affinity introduced by protein mutations. (C) The pipeline of the proposed SPLDExtraTrees method. SPLDExtraTrees gradually incorporates both easy (with the small loss values) and diversity samples (from different protein families) into learning, and then iterates between weight recalculating and model updating. A dotted block denotes a group with given $\lambda$ and $\gamma$, colored box denotes a sample, and the samples in each group are sorted according to the loss values (color from dark to light).}
\label{Figure_1}
\end{figure*}

In this section, we first present the dataset used for training and testing machine learning methods. The calculation of input features associated with affinity changes upon protein mutation and feature selection will then be introduced. After that, we present the objective function of the proposed SPLDExtraTrees method as well as an efficient algorithm for realizing SPLDExtraTrees. The pipeline of the proposed method is summarized in Figure~\ref{Figure_1}.

\subsection{Datasets}
\begin{figure}[]
\centering
\includegraphics[width=0.5\textwidth]{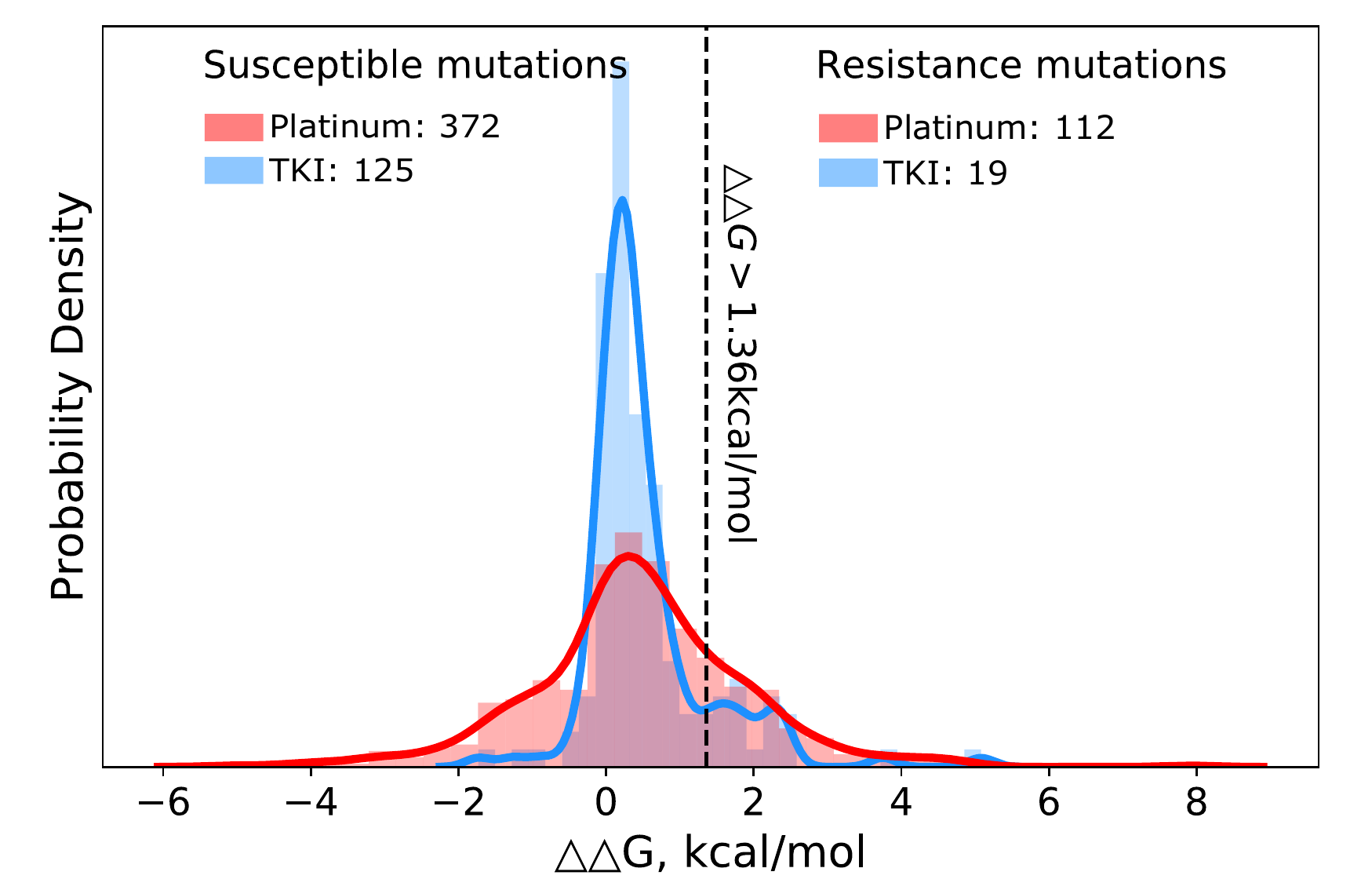}
\caption{Distribution of the experimental $\Delta\Delta$G values of the Platinum and TKI datasets. The line at $\Delta\Delta G= 1.36$ kcal/mol separates mutations defined as resistant from susceptible.}
\label{Figure1}
\end{figure}

The machine learning methods are trained on a subset of the Platinum dataset \cite{pires2015platinum} and tested on the TKI dataset \cite{hauser2018predicting}. Following the data preprocessing process outlined in \cite{aldeghi2019predicting}, we exclude some samples from the whole training dataset if they: (1) do not have a point-mutated correspondent in the dataset; (2) contain ``broken ligand structures''; (3) contain ligands that dissolve poorly. Finally, the training dataset contains 484 point-mutations with associated affinity change values ($\Delta\Delta$G) across 84 proteins and 143 ligands. For the test dataset, there are 144 point-mutation samples with TKI affinity changes ($\Delta\Delta$G) across 8 TKIs caused by 31 Abl mutations. Table \ref{Table1} summarizes the TKI dataset for 144 Abl kinase mutations and 8 TKIs used in this work. Furthermore, Figure~\ref{Figure1} shows the distribution of experimental $\Delta\Delta$G values for the Platinum and TKI datasets.

\begin{table}[]
\centering
\caption{Detailed information of the TKI dataset used in the experiments.}
\begin{threeparttable} 
\setlength{\tabcolsep}{3mm}{
\begin{tabular}{ccccc}
\hline
\textbf{TKI} & \textbf{N$_{mut}$}\tnote{1} & \textbf{Resistant}\tnote{2} & \textbf{Susceptible}\tnote{2} & \textbf{PDB} \\ \hline
Axitinib  & 26   & 0         & 26          & 4WA9 \\ 
Bosutinib & 21   & 4         & 17          & 3UE4 \\ 
Dasatinib & 21   & 5         & 16          & 4XEY \\ 
Imatinib  & 21   & 5         & 16          & 1OPJ \\ 
Nilotinib & 21   & 4         & 17          & 3CS9 \\ 
Ponatinib & 21   & 0         & 21          & 3OXZ \\ 
Erlotinib & 7    & 1         & 6           & Dock to 3ue4\tnote{3} \\ 
Gefitinib & 6    & 0         & 6           & Dock to 3ue4\tnote{3} \\ \hline
Total     & 144  & 19        & 125         &      \\ \hline
\end{tabular}}
\begin{tablenotes}
\footnotesize
\item[$^{1}$] N$_{mut}$ represents the total number of Abl kinase mutants.
\item[$^{2}$] Number of resistant, susceptible mutants using 10-fold affinity change threshold, corresponding to a binding free energy change of 1.36 kcal/mol.
\item[$^{3}$] PDB Source PDB ID, or Dock to 3ue4, which used 3ue4 as the receptor for Glide-SP docking inhibitors without co-crystal structure \cite{hauser2018predicting}.\vspace*{6pt}
\end{tablenotes}
\end{threeparttable}
\label{Table1}
\end{table}

\subsection{Feature Calculation}
We calculate a total of 146 features that bear potentially useful information for predicting affinity changes introduced by protein mutations. 

Overall, 127 features are calculated, including crystallographic protein-ligand structures, physicochemical properties of ligands and residues, and the fast empirical scoring function for ligand-target binding. These feature selections are consistent with the work in \cite{aldeghi2019predicting}. Specifically, 18 ligand properties (e.g. logP, molecular weight, and apolar surface area) are calculated using RDkit version 2018.09.1. 13 residue mutation properties are calculated using precomputed properties for each amino acid (e.g. change in hydropathy index, flexibility, volume and so on). Among these, there are also the changes in folding free energy upon mutations that predicted by FoldX v4 \cite{schymkowitz2005foldx}. 21 features describing the mutation environment (e.g. radial count of protein atoms near the mutation, number of polar/apolar/charged residues in the binding pocket) are calculated using Biopython version 1.73. 6 protein-ligand interaction features are calculated using the Protein-Ligand Interaction Profiler (PLIP) \cite{salentin2015plip}. 59 Vina features are calculated with the modified AutoDock Vina program \cite{trott2010autodock} provided with DeltaVINA \cite{wang2017improving}. We also use DeltaVINA to calculate 10 pharmacophore-based solvent accessible surface area (SASA) features. 

In addition, we also integrate energy features associated with empirical and statistical potentials, aiming to capture more sophisticated protein-ligand complex information than that encoded in the traditional features \cite{alford2017rosetta}. We incorporate 19 Rosetta energy features, which are computed using the Rosetta full-atom force field \cite{alford2017rosetta}. Specifically, the ligands are parametrized using Rosetta’s molfile\_to\_params.py script. The interface between the ligand and the protein is then minimized with RosettaScript. The XML description of the complete process is available from the RosettaScripts official website (https://new.rosettacommons.org/docs/latest). A total of 10 complexes are generated for each protein-ligand complex, and the full-atom energies for the structure with the lowest energy score are extracted. The 19 full-atom energy features in REF15 are incorporated into the feature set. 

Finally, the differences between the wild-type and mutant feature values are calculated and fed into machine-learning models for predicting the $\Delta\Delta$G value for the mutation-induced difference in the binding free energy. Figures S1-S6 in the supplementary material show the distribution of each selected features on the Platinum and TKI datasets.

\subsection{Feature Selection}
Feature selection aims to select a subset of the initial features by removing those that possess less predictive power. Generally, feature selection methods can be roughly divided into three categories: filter, wrapper, and embedded methods \cite{liang2013sparse}. Filter methods evaluate a feature based on discriminative power without considering its correlations with other features \cite{dudoit2002comparison, li2004comparative, lee2005extensive, ding2005minimum}. Wrapper methods utilize a particular machine learning method as feature evaluation metric, select a subset of features according to the estimated prediction error, and build the final learning machine \cite{monari2000withdrawing, rivals2003mlps, golub1999molecular}. However, the wrapper methods greatly require extensive computational time. For the embedded method, it performs the feature selection as part of the learning procedure, which is more computationally efficient than the wrapper method \cite{liang2013sparse, yang2018robust}.

In this work, feature selection is performed with the SelectFromModel function using the sklearn library, which is an embedded feature selection method. When the tree-based model is coupled with the SelectFromModel meta-transformer, feature importance can be calculated and used to discard irrelevant features. We allow the selection of any number of features, which minimizes the mean-squared error of $k$-fold ($k=10$) cross-validation on the Platinum dataset. The distribution of each selected features on the Platinum and TKI datasets is shown in Figure S7 in the supplementary material.

\subsection{Proposed Method}
We show that machine learning can achieve state-of-the-art results in predicting ligand binding affinity changes upon protein mutation if appropriate learning strategy is adopted. The proposed SPLDExtraTrees method is built on top of the basis of extremely randomized regression trees (ExtraTrees) \cite{aldeghi2019predicting}. Instead of learning from all samples simultaneously, we introduce self-paced learning (SPL) strategy that learns in a meaningful order, from easy samples to hard ones \cite{bengio2009curriculum, kumar2010self} according to the definition of easiness provided in the subsequent discussion. In the sample selection process, we incorporate protein family information so that the model can obtain more comprehensive knowledge during the learning process and achieve better generalization performance.

\subsubsection{Model Formulation}
Theoretically, given the training dataset $\mathcal{D}=\left\{\left(\bm{x}_{i}, y_{i}\right)\right\}_{i=1}^{n}$, where $\bm{x}_{i} \in \mathbb{R}^{m}$ represents the $i$-th input sample, and $y_{i} \in \mathbb{R}$ represents the corresponding experimental $\Delta\Delta$G value. The model learns a mapping from input to output through the decision function $f_{\bm{\beta}}:X \rightarrow Y$, where $\bm{\beta}$ is the model parameter inside the decision function. Let $L\left(y_{i}, f\left(\bm{x}_{i}, \bm{\beta}\right)\right)$ denote the loss function which calculates the cost between the experimental $\Delta\Delta$G value $y_{i}$ and the estimated $\Delta\Delta$G value $f\left(\bm{x}_{i}, \bm{\beta}\right)$. Here, we adopt the most commonly used Root Mean Square Error (RMSE) as the loss function $L$ and apply ExtraTrees as the decision function $f$. The objective in a conventional machine learning model can be expressed as:
\begin{equation}
\label{Eq_01}
\min _{\bm{\beta}} \mathbf{E}(\bm{\beta})=\sum_{i=1}^{n}  L\left(y_{i}, f\left(\bm{x}_{i}, \bm{\beta}\right)\right)
\end{equation}

To improve the generalization ability in predicting binding affinity changes for protein mutations, we incorporate self-paced regularization term into the learning objective to adaptively learn the model in a meaningful order. Specifically, we define a latent weight variable $\bm{v}=\left[v_{1}, \cdots, v_{n}\right]$. The purpose of the SPL model is to jointly learn the model parameter $\bm{\beta}$ and the latent weight variable $\bm{v}$ by minimizing:
\begin{equation}
\label{Eq_02}
\min _{\bm{\beta}, \bm{v} \in[0,1]^{n}} \mathbf{E}(\bm{\beta}, \bm{v}; \lambda)=\sum_{i=1}^{n} v_{i} L\left(y_{i}, f\left(\bm{x}_{i}, \bm{\beta}\right)\right)-\lambda \sum_{i=1}^{n} v_{i}
\end{equation}
where $\lambda$ denotes the age parameter for adjusting the learning space. Eq.~(\ref{Eq_02}) is to minimize the weighted training loss together with the negative $l_{1}$-norm regularizer $-\|\bm{v}\|_{1}=-\sum_{i=1}^{n} v_{i}$ (since $\left.v_{i} \geq 0\right)$). 

In the standard SPL model, samples are selected solely in terms of ``easiness''. In practice, however, diversity is an important aspect of learning that should also be considered. For instance, there are large set of protein families with differing binding properties \cite{das2015diversity}. Typically, the correlation between data points is higher when the samples belong to the same protein family as opposed to different ones. Therefore, we encourage the consideration of protein diversity in sample selection during self-paced training, which implies that the selected samples should be less similar. To achieve this, we embed a diversity regularization term \cite{jiang2014self} into Eq.~(\ref{Eq_02}).

Formally, assume training samples $\bm{X}=(\bm{x}_{1}, \dots, \bm{x}_{n}) \in \mathbb{R}^{m \times n}$ are partitioned into $b$ protein groups: $\bm{X}^{(1)}, \dots, \bm{X}^{(b)}$, where columns of $\bm{X}^{(j)} \in \mathbb{R}^{m \times n_{j}}$ correspond to the samples in the $j$-th group, $n_{j}$ is the number of samples under the $j$-th group, and $\sum_{j=1}^{b} n_{j}=n$. Accordingly, denote the weight vector as $\bm{v} =\left[\bm{v}^{(1)}, \dots, \bm{v}^{(b)}\right]$, where $\bm{v}^{(j)} = \left(v_{1}^{(j)}, \dots, v_{n_{j}}^{(j)}\right)^{T} \in[0,1]^{n_{j}}$. On the one hand, we expect that the model needs to assign a non-zero weight to $\bm{v}$, and on the other hand, it needs to distribute the non-zero weight to more protein groups $\bm{v}^{(j)}$ to increase diversity. Both of these requirements can be simultaneously achieved with the following optimization model:
\begin{equation}
\label{Eq_04}
\min _{\bm{\beta}, \bm{v} \in[0,1]^{n}} \mathbb{E}(\bm{\beta}, \bm{v} ; \lambda, \gamma)=\sum_{i=1}^{n} v_{i} L\left(y_{i}, f\left(\bm{x}_{i}, \bm{\beta}\right)\right)-\lambda \sum_{i=1}^{n} v_{i}-\gamma\|\bm{v}\|_{2,1}
\end{equation}
where $\lambda$ and $\gamma$ are the parameters imposed on the easiness term (the negative $l_{1}$-norm: $-\|\bm{v}\|_{1}$) and the diversity term (the negative $l_{2,1}$-norm: $-\|\bm{v}\|_{2,1}$), respectively. Furthermore, the diversity term can be expressed as:
\begin{equation}
\label{Eq_05}
-\|\bm{v}\|_{2,1}=-\sum_{j=1}^{b}\left\|\bm{v}^{(j)}\right\|_{2}
\end{equation}
As shown in Eq.~(\ref{Eq_04}), this SP-regularizer consists of two components. One is the negative $l_{1}$-norm inherited from the conventional SPL, which favors selecting easy over complex samples. The other one is the negative $l_{2,1}$-norm, which tends to select diverse samples scattered in different protein groups. It is well known that the $l_{2,1}$-norm leads to the group sparsity \cite{yuan2006model}, that is, the non-zero entries of $\bm{v}$ are likely to be concentrated in a few groups. On the contrary, the negative $l_{2,1}$-norm should have a counter-effect to group sparsity. Therefore, this self-paced learning with diversity (SPLD) regularization term selects not only easy samples but also diverse samples that are sufficiently dissimilar from what has already been learned.

In this paper, we incorporate extremely randomized regression trees (ExtraTrees) with SPLD. Each node of the regression trees is trained with easy and diverse samples selected in a self-paced way. Compared with the conventional ExtraTrees method, SPLDExtraTrees is expected to better avoid overfitting and manifest a superior generalization capability.

\subsubsection{Model Algorithm}
We adopt an alternating optimization strategy to solve Eq.~(\ref{Eq_04}), the details are listed in Algorithm~\ref{alg_02}. A challenge is that optimizing $\bm{v}$ with a fixed $\bm{\beta}$ becomes a non-convex problem. Jiang \emph{et al.} \cite{jiang2014self} proposed a simple yet effective algorithm to obtain the global optimum solution of the model, as listed in Algorithm~\ref{alg_01}. According to Step 5 of Algorithm~\ref{alg_01}, a sample whose loss is smaller than the threshold $\lambda+\gamma \frac{1}{\sqrt{i}+\sqrt{i-1}}$ will be selected in the learning process, where $i$ represents the sample's rank with respect to its loss value within its group. Since the threshold value gradually decreases with the increase of rank $i$, Step 5 penalizes samples selected monotonically from the same group. Therefore, when Algorithm~\ref{alg_01} obtains the optimal $\bm{v}^{*}$, the alternative optimization strategy algorithm can be easily applied to solve Eq.~(\ref{Eq_04}). Following the work in \cite{kumar2010self}, we initialize $\bm{v}$ by setting $v_{i}=1$ to randomly selected samples. Following SPL \cite{kumar2010self}, the self-paced parameters are updated by absolute values of $\mu_{1}$, $\mu_{2}$ ($\mu_{1}$, $\mu_{2}$ $\geq$ 1) in Step 5 of Algorithm~\ref{alg_02} at the end of every iteration. According to the work in \cite{jiang2014self}, it proposes that the algorithm seems more robust by first sorting samples in ascending order of their losses, and then setting the $\lambda$ and $\gamma$ according to the statistics collected from the ranked samples. In this paper, we adopt this mechanism to update the parameters.

\begin{algorithm}[t]  
  \caption{Algorithm for realizing SPLDExtraTrees.}
  \label{alg_02}
  \begin{algorithmic}[1]  
    \Require  
      Input dataset $\mathcal{D}$, self-pace parameters $\lambda$, $\gamma$, and max iterations $T$.
    \Ensure
      Model parameter $\bm{\beta}$.
    \State Initialize $\bm{v}^{*}$, $\lambda$, $\gamma$;
    \While{not converge || $t \leq T$}
        \State Update $\bm{\beta}^{*}=\arg \min _{\bm{\beta}} \mathbb{E}\left(\bm{\beta}, \bm{v}^{*} ; \lambda, \gamma\right)$; // train a regression model
        \State Update $\bm{v}^{*}=\arg \min _{\bm{v}} \mathbb{E}\left(\bm{\beta}^{*}, \bm{v} ; \lambda, \gamma\right)$ using Algorithm~\ref{alg_01}; // select easy and diverse samples
        \State $\lambda \leftarrow \mu_{1} \lambda$ ; $\gamma \leftarrow \mu_{2} \gamma$; // update the learning pace
    \EndWhile
    \State \textbf{return} $\bm{\beta}=\bm{\beta}^{*}$
  \end{algorithmic}
\end{algorithm}

\begin{algorithm}[t]  
  \caption{Algorithm for Solving $\min _{\bm{\bm{v}}} \mathbb{E}(\bm{\beta}, \bm{v} ; \lambda, \gamma)$.}  
  \label{alg_01}  
  \begin{algorithmic}[1]  
    \Require  
      Input dataset with $b$ groups: $\bm{X}^{(1)}, \cdots, \bm{X}^{(b)}$, $\bm{\beta}$, $\lambda$ and $\gamma$.
    \Ensure  
      The global optimum solution $\bm{v}=\left(\bm{v}^{(1)}, \cdots, \bm{v}^{(b)}\right)$ of $\min _{\bm{\bm{v}}} \mathbb{E}(\bm{\beta}, \bm{v} ; \lambda, \gamma)$.
    \For{$i=1$ to $b$} // for each protein group
        \State Sort the samples $\bm{X}^{(j)}$ as $\left(\bm{x}_{1}^{(j)}, \cdots, \bm{x}_{n_{j}}^{(j)}\right)$ in ascending order of loss values $L$;
        \State Accordingly, denote the experimental $\Delta \Delta$G values and weights of $\bm{X}^{(j)}$ as $\left(y_{1}^{(j)}, \cdots, y_{n_{j}}^{(j)}\right)$ and $\left(v_{1}^{(j)}, \cdots, v_{n_{j}}^{(j)}\right)$;
        \For{$i=1$ to $n_{j}$}
            \If{$L\left(y_{i}^{(j)}, f\left(\boldsymbol{x}_{i}^{(j)}, \bm{w}\right)\right)<\lambda+\gamma \frac{1}{\sqrt{i}+\sqrt{i-1}}$} $v_{i}^{(j)}=1$; // select this sample
            \Else{ $v_{i}^{(j)}=0$ } // not select this sample
            \EndIf 
        \EndFor
    \EndFor  
  \end{algorithmic}  
\end{algorithm}

\section{Results and Discussion}
In this section, we compare the proposed method, SPLDExtraTrees, with three different types of computational methods: the molecular dynamics methods, Rosetta, and the machine learning methods. To evaluate the capability of the proposed method for predicting drug response to tyrosine kinase site-mutations, we test the effectiveness of the model on the TKI dataset \cite{hauser2018predicting, aldeghi2019predicting}.

\subsection{Experimental Setup}
We conduct experiments on the TKI dataset under three scenarios. The first scenario is to train the machine learning methods on the Platinum dataset, and then test it on the TKI dataset. In this scenario, tyrosine kinase is not present in the training dataset, such that we can evaluate whether the model could extrapolate to this protein target. The second scenario is to train the model on the Platinum dataset and a small portion of the TKI dataset, and test it on the rest of the TKI dataset. To do this, we randomly select 2 samples from each inhibitor in the TKI dataset and feed them into the training pool along with the Platinum dataset. Compared with Scenario 1, we evaluate whether the model could effectively improve its ability to predict affinity changes in Abl mutants by learning from a small amount of tyrosine kinase information.

The third scenario is to train and test the model on the TKI dataset. We apply 8-fold nested cross-validation, as set up in work \cite{aldeghi2019predicting}. In each iteration, the model selects a subset of the feature set through 7-fold cross-validation with 7 inhibitors, and test on the remaining one. This process is performed 8 times to obtain the prediction results for all entries on the TKI dataset. In this scenario, the model is trained on tyrosine kinase mutations only, such that we can evaluate the interpolating capability of the model.

\subsection{Models in Comparison}
We compare SPLDExtraTrees with three types of methods. The first is molecular dynamics (MD) simulations, using the Amber99sb*-ILDN force field with two different simulation protocols A99 and A99$l$, respectively \cite{wang2004development, hornak2006comparison, best2009optimized, lindorff2010improved}. For simplicity, we show only the results obtained with the best-performing Amber force fields, which are reported by \cite{aldeghi2019predicting}. The second is Rosetta, a modeling program that uses mixed physics- and knowledge-based potentials. The REF15 scoring function, among them, achieves the best published results on this task \cite{aldeghi2019predicting}. The third fold is machine learning methods. We apply ExtraTrees \cite{aldeghi2019predicting} and ExtraTrees with self-paced learning strategy (SPLExtraTrees) as the competing methods.

\subsection{Evaluation Metrics}
Performance are evaluated by the Root Mean Square Error (RMSE) between the experimentally measured and calculated $\Delta\Delta$G values, the Pearson correlation coefficient (Pears), and the area under the precision recall curve (AUPRC). The latter measures the classification ability to classify mutations as resistant or susceptible. Particularly, precision is calculated by dividing the number of the true positive resistant mutations by the total number of predicted resistant mutations, and recall is calculated by dividing the number of the true positive resistant mutations by the total number of true resistant mutations. Consistent with previous work \cite{hauser2018predicting, aldeghi2019predicting}, resistant mutations are defined as the affinity changes for mutants by least 10-fold, i.e., $\Delta\Delta$G$_{exp} \textgreater 1.36$ kcal/mol.

\begin{table*}[t]
\centering
\scriptsize
\caption{Summary of the computational methods used, their calculation costs and performance. Mean prediction performance $x^{upper}_{lower}$ over 20 repetitions are reported. The best results are highlighted in \textbf{bold}.}
\renewcommand{\arraystretch}{2}
\begin{threeparttable}
\setlength{\tabcolsep}{3mm}{
\begin{tabular}{|c|c|c|c|c|c|c|c|c|}
\hline
\multirow{2}{*}{\textbf{Abbreviation}} & \multicolumn{2}{c|}{\multirow{2}{*}{\textbf{Method}}} & \multirow{2}{*}{\textbf{\makecell[c]{Force field \\or scoring function}}} & \multicolumn{2}{c|}{\textbf{\makecell[c]{Approximate cost \\ per $\Delta\Delta$G estimate}}} & \multicolumn{3}{c|}{\textbf{Performance}}                      \\ \cline{5-9} 
& \multicolumn{2}{c|}{} &  & \textbf{Hardware}  & \textbf{\makecell[c]{Compute \\ hours}}     & \textbf{\makecell[c]{RMSE \\ (kcal/mol)}} & \textbf{Pearson} & \textbf{AUPRC} \\ \hline
A99$^{1}$                           & \multicolumn{2}{c|}{Molecular Dynamics}      & \makecell[c]{Amber99sb*-ILDN \\ and   GAFF v2.1}                  & \makecell[c]{10 CPU cores \\ and 1 GPU}         & 59    & $0.91^{1.05}_{0.77}$   & $0.44^{0.59}_{0.20}$    & \bm{$0.56^{0.77}_{0.32}$}  \\ \hline
A99$l$$^{1}$                           & \multicolumn{2}{c|}{Molecular Dynamics}      & \makecell[c]{Amber99sb*-ILDN \\ and   GAFF v2.1}                  & \makecell[c]{10 CPU cores \\ and 1 GPU}   & 98   & $0.91^{1.09}_{0.74}$    & $0.42^{0.59}_{0.20}$    & $0.51^{0.75}_{0.27}$  \\ \hline
REF15$^{2}$    & \multicolumn{2}{c|}{Rosetta}   & REF15    & 1 CPU core        & 32                 & $0.72^{0.83}_{0.60}$   & $0.67^{0.81}_{0.45}$    & $0.53^{0.74}_{0.29}$  \\ \hline
ExtraTrees$^{*3}$                     & ML & \multirow{3}{*}{Scenario 1} & n/a                                              & 1 CPU core                       & 0.02                  & $0.87^{1.06}_{0.68}$              & $0.12^{0.29}_{-0.04}$    & $0.20^{0.39}_{0.10}$  \\ \cline{1-2} \cline{4-9} 
SPLExtraTrees          & ML &                             & n/a                                              & 1 CPU core                       & 0.02                  & $0.75^{0.77}_{0.75}$              & $0.50^{0.54}_{0.38}$    & $0.48^{0.52}_{0.34}$  \\ \cline{1-2} \cline{4-9} 
SPLDExtraTrees          & ML &                             & n/a                                              & 1 CPU core                       & 0.02                  & $0.73^{0.74}_{0.72}$              & $0.54^{0.56}_{0.47}$    & $0.50^{0.52}_{0.43}$  \\ \hline
ExtraTrees           & ML &        \multirow{3}{*}{Scenario 2}                      & n/a                                              & 1 CPU core                       & 0.02                  & $0.81^{0.89}_{0.66}$              & $0.34^{0.54}_{0.22}$    & $0.35^{0.47}_{0.22}$  \\ \cline{1-2} \cline{4-9} 
SPLExtraTrees           & ML &                             & n/a                                              & 1 CPU core                       & 0.02                  & $0.73^{0.80}_{0.53}$              & $0.53^{0.65}_{0.38}$    & $0.46^{0.57}_{0.35}$  \\ \cline{1-2} \cline{4-9} 
SPLDExtraTrees          & ML &                             & n/a                                              & 1 CPU core                       & 0.02                  & $0.70^{0.76}_{0.57}$              & $0.60^{0.68}_{0.49}$    & $0.55^{0.72}_{0.42}$  \\ \hline
ExtraTrees$^{*3}$                     & ML & \multirow{3}{*}{Scenario 3} & n/a                                              & 1 CPU core                       & 0.02                  & $0.68^{0.80}_{0.55}$              & $0.57^{0.72}_{0. 34}$    & $0.47^{0.68}_{0.25}$  \\ \cline{1-2} \cline{4-9} 
SPLExtraTrees           & ML &                             & n/a                                              & 1 CPU core                       & 0.02                  & $0.59^{0.61}_{0.58}$              & $0.72^{0.73}_{0.70}$    & \bm{$0.56^{0.57}_{0.51}$}  \\ \cline{1-2} \cline{4-9} 
SPLDExtraTrees          & ML &                             & n/a                                              & 1 CPU core                       & 0.02                  & \bm{$0.58^{0.59}_{0.57}$}              & \bm{$0.74^{0.75}_{0.72}$}    & \bm{$0.56^{0.60}_{0.52}$}  \\ \hline
\end{tabular}}
\begin{tablenotes}
\scriptsize
\item[1] Data for the molecular dynamic simulations with the A99 and A99$l$ force field are obtained from the work in \cite{aldeghi2019predicting}.
\item[2] Data for the Rosetta REF15 scoring function are obtained from the work in \cite{aldeghi2019predicting}.
\item[3] Data for the ExtraTrees$^{*}$ are obtained from the work in \cite{aldeghi2019predicting}.
\end{tablenotes}
\end{threeparttable}
\label{Table2}
\end{table*}

\subsection{Results}

\begin{figure*}[t]
\centering
\includegraphics[width=1\textwidth]{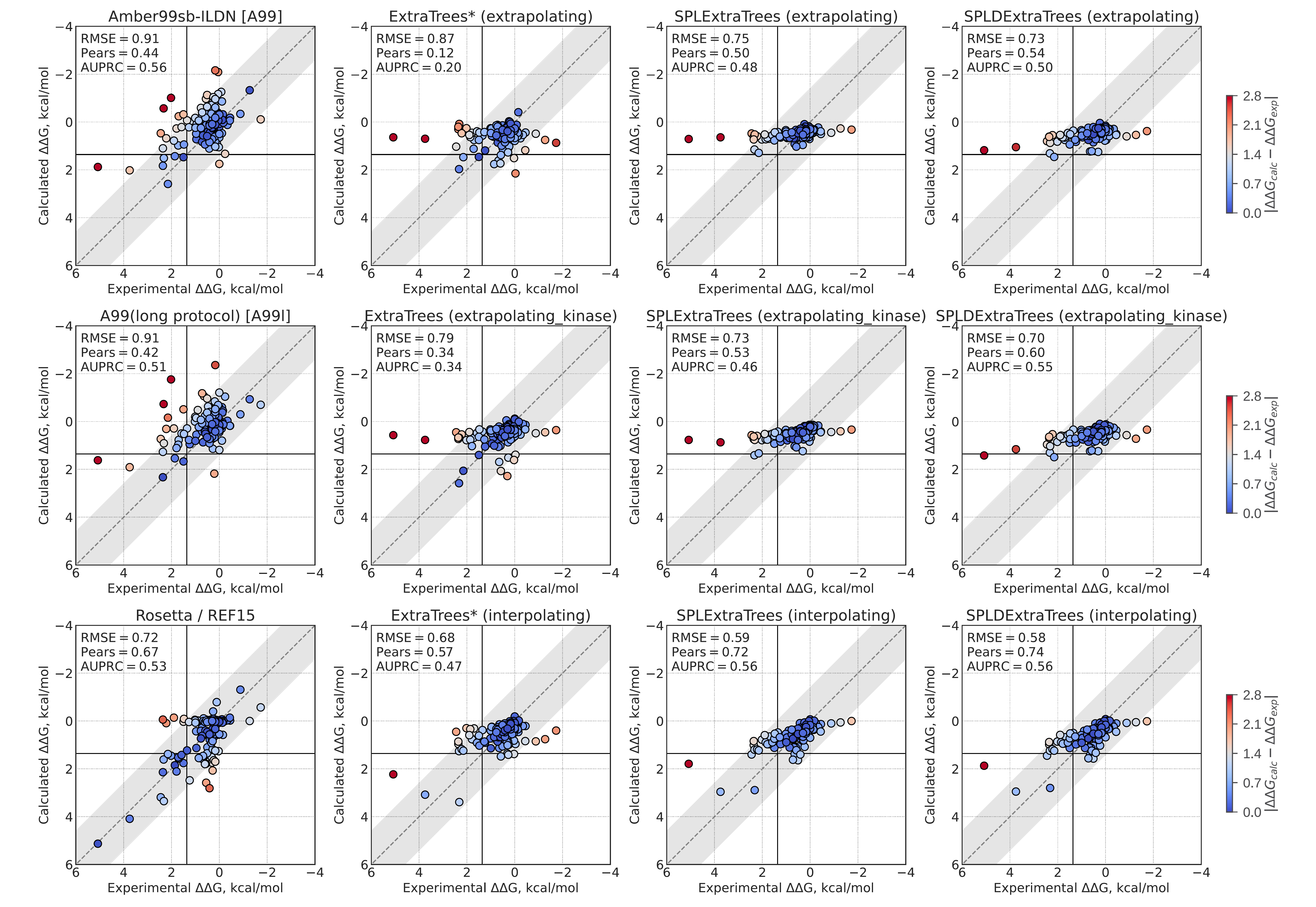}
\caption{Scatter plots of the experimental versus calculated $\Delta\Delta$G values. $x$-axis denotes the experimental $\Delta\Delta$G value (kcal/mol). $y$-axis denotes the calculated $\Delta\Delta$G value (kcal/mol). Each $\Delta\Delta$G estimate is color-coded according to its absolute error w.r.t. the experimental $\Delta\Delta$G value; at 300 $K$, the 1.4 kcal/mol error corresponds to a ~10-fold error in the $K_d$ change, and the 2.8 kcal/mol error corresponds to a ~100-fold error in the $K_d$ change. Extrapolating represents the model is trained on the Platinum dataset and tested on the TKI dataset, i.e., Scenario 1. Extrapolating\_kinase represents the model is trained on Scenario 2. Interpolating represents the model is trained and tested on the TKI dataset, i.e., Scenario 3.}
\label{Figure2}
\end{figure*}

Table~\ref{Table2} summarizes the performance of the $\Delta\Delta$G estimates across all competing methods on the TKI dataset, including computational costs (see Table S1 in the supplementary material for more details). In terms of computational costs, molecular dynamics methods are the most expensive to obtain the accurate $\Delta\Delta$G estimates. On the contrary, the machine learning methods can obtain $\Delta\Delta$G estimates from a few seconds to minutes on a single CPU core, as long as the necessary structure-based features are calculated in advance. In terms of binding affinity changes prediction and resistant mutation identification, we propose three scenarios to evaluate the extrapolating and interpolating capability of the proposed method. As we can see in Table~\ref{Table2}, SPLDExtraTrees outperforms ExtraTrees, SPLExtraTrees and MD under all three scenarios, and it is comparable with Rosetta's results for Scenarios 2 and 3, except Pearson in Scenario 2.

For the first scenario, we trained the machine learning methods on the Platinum dataset and tested them on the TKI dataset to evaluate the model's extrapolating capability. In this scenario, the Platinum dataset was pre-filtered to excluded samples that belonged to the same protein family with Abl kinases in order to assess the generalization capacity of the models. Figure~\ref{Figure2} plots the scatter plots of the experimental versus calculated $\Delta\Delta$G values. As shown in the upper panel, the SPL and SPLD strategy helped improve the robustness of the ExtraTrees in predicting $\Delta\Delta$G. SPL alone could enhance the performance of ExtraTrees$^*$ (RMSE = 0.87 kcal/mol, Pearson = 0.12, and AUPRC = 0.20) by a large margin, which yielded more accurate predictions on $\Delta\Delta$G with lower absolute errors (RMSE = 0.75 kcal/mol), stronger correlation (Pearson = 0.50), and better classification performance (AUPRC = 0.48). The protein family information was incorporated in the SPLDExtraTrees, which further improved the prediction performances (RMSE = 0.73 kcal/mol, Pearson = 0.54, and AUPRC = 0.50). Meanwhile, the additional physical and structural features (i.e., obtained by the Rosetta REF15 scoring function) also provided useful information and contributed to prediction improvement (see Figure S8 and Table S1 in the supplementary material). We observed that the range of $\Delta\Delta$G predicted by SPLExtraTrees and SPLDExtraTrees is narrow, from 0.04 to 1.46, which might still be a challenge (requiring further analyses in the future). Interestingly, SPLDExtraTrees outperformed molecular dynamics methods (A99 and A99$l$) by a considerable margin, which demonstrated its generalization ability for the kinase, as an example. However, SPLDExtraTrees underperformed Rosetta (REF15) (RMSE = 0.72, Pearson = 0.67, and AUPRC = 0.53) slightly. This might result from the absence of samples in kinase family in the training set, which are more closely related to the Abl proteins.

Therefore, in the second scenario, a small part of the TKI dataset along with the Platinum dataset was used to train the models, and the rest of the TKI dataset was used for testing. We assumed that the model could improve its ability to predict $\Delta\Delta$G by learning from a small amount of tyrosine kinase information. As shown in the middle panel of Figure~\ref{Figure2}, all three methods got a higher prediction accuracy than that in the scenario 1. Specifically, we could see a larger improvement between SPLExtraTrees and SPLDExtraTrees when samples of the same protein family were presented in the training set. Even with this small amount of kinase samples, SPLDExtraTrees obtained good estimates (RMSE = 0.70 kcal/mol and AUPRC = 0.55) that were comparable with Rosetta calculations (RMSE =0.72 kcal/mol and AUPRC = 0.53), except Pearson metric (Table~\ref{Table2}). 

For the third scenario, the machine learning methods were trained and tested on the TKI dataset such that we could evaluate the interpolative capability of the model. As shown in the bottom panel of Figure~\ref{Figure2}, SPLDExtraTrees got the best performance across the benchmark. Pearson and AUPRC achieved by SPLDExtraTrees were increased by approximately more than 0.18 and 0.15 in comparison to those for ExtraTrees$^*$, respectively. Futhermore, the additional pyhsical and strutral features could also provide the valuable information to improve the predictive performance (see Figure S9 and Table S1 in the supplementary material for detail). However, Because there was only one family in both training and test sets, SPLDExtraTrees did not get extra information from the protein family, and got comparable results with SPLExtraTrees. Moreover, SPLDExtraTrees attained more accurate $\Delta\Delta$G estimates with lower absolute errors (RMSE = 0.58 kcal/mol), higher correlation (Pearson = 0.74), and better classification performance (AUPRC = 0.56) compared with Rosetta (RMSE = 0.72 kcal/mol, Pearson = 0.67, and AUPRC = 0.53). It implies that the proposed method can obtain state-of-the-art performance of $\Delta\Delta$G estimates and could discriminate well between resistant and susceptible mutations when it is trained on the relevant data (i.e., tyrosine kinase).

\begin{figure*}[t]
\centering
\includegraphics[width=0.9\textwidth]{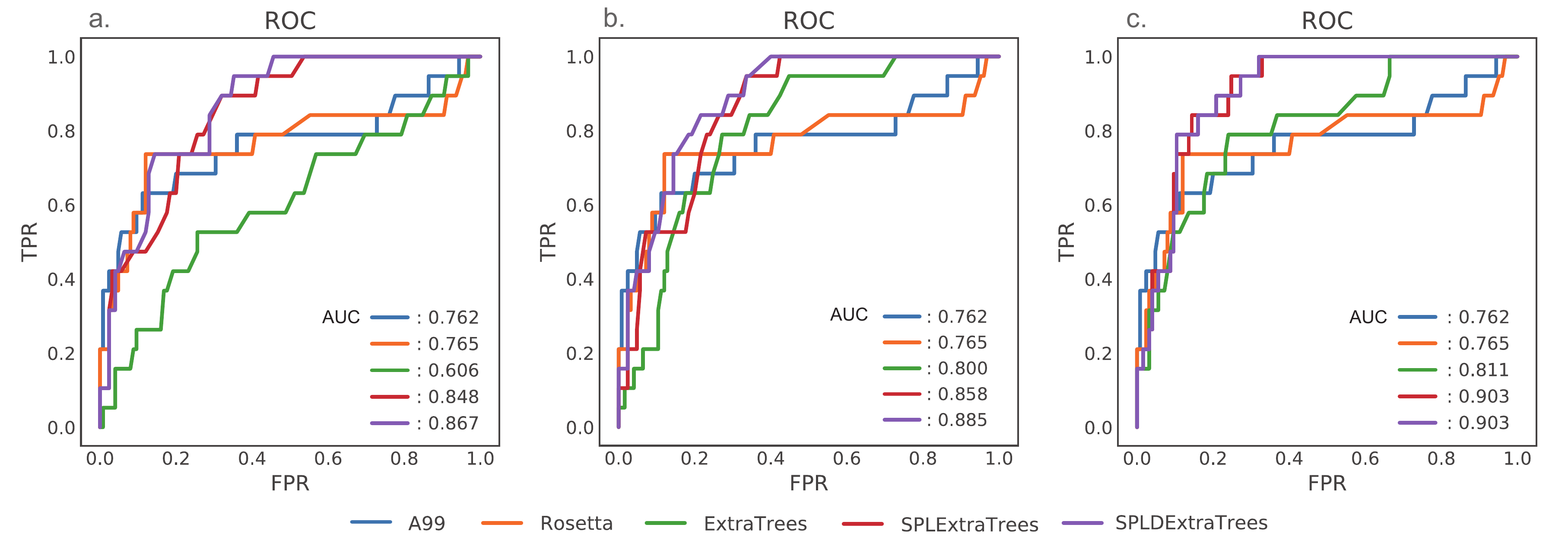}
\caption{All competing methods validation using Receiver Operating Characteristic (ROC) curves. The $x$-axis represents the False Positive Rate (FPR), and the $y$-axis represents the True Positive Rate (TPR). (a) The machine learning methods are trained in the extrapolation case, i.e., Scenario 1. (b) The machine learning methods are trained in the extrapolation with few kinase samples, i.e., Scenario 2. (c) The machine learning methods are trained in the interpolation case, i.e., Scenario 3.}
\label{Figure3}
\end{figure*}

\begin{figure*}[t]
\centering
\includegraphics[width=0.9\textwidth]{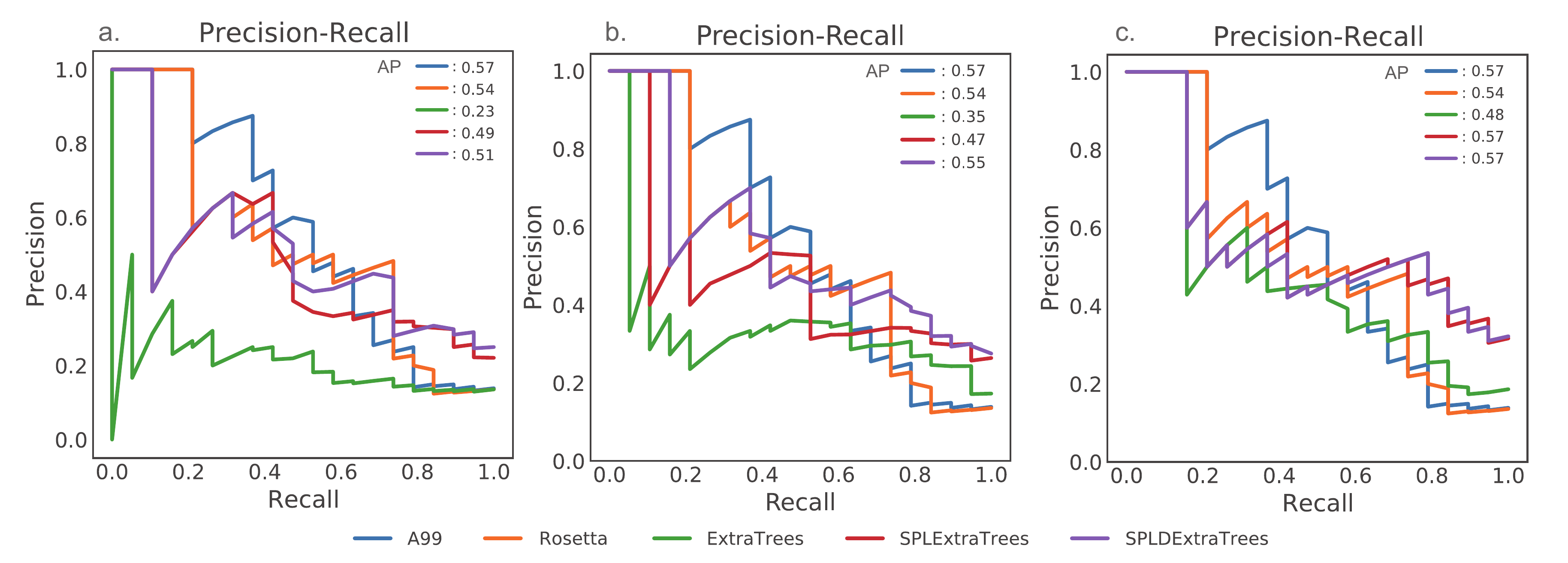}
\caption{All competing methods validation using Precision Recall Curve (PRC). The $x$-axis represents Recall, and the $y$-axis represents Precision. (a) The machine learning methods are trained in the extrapolation case. (b) The machine learning methods are trained in the extrapolation with few kinase samples. (c) The machine learning methods are trained in the interpolation case. }
\label{Figure4}
\end{figure*}

To further illustrate the capability of the proposed method to accurately identify resistant mutations, we plot the Receiver Operating Characteristic (ROC) curve. As shown in Figure~\ref{Figure3}, we can clearly see that the proposed method outperforms all the competing methods by a large margin under all three scenarios. In the extrapolating case, as shown in Figure~\ref{Figure3} (a), SPLDExtraTrees obtained AUC with 0.867, and attained more than 10\% improvement compared with molecular dynamics A99 and Rosetta (REF15). AUC achieved by SPLDExtraTrees is increased by approximately more than 14\% and 13\% gain compared with MD (A99) and Rosetta (REF15) in the extrapolating case, respectively, as shown in Figure~\ref{Figure3} (c). However, most protein mutations are resistant rather than susceptible. Therefore, Precision Recall Curve (PRC) is more informative than the ROC curve when evaluating the models on such an imbalanced data distribution. Figure~\ref{Figure4} plots the precision recall curve calculated on the TKI dataset for both extrapolation and interpolation scenarios. As shown in Figure~\ref{Figure4} (a), although the proposed method outperformed ExtraTrees$^*$ by a large margin, it performed slightly inferior to molecular dynamics simulations (A99) and Rosetta (REF15) in the case of extrapolation. The average presicion (AP) of SPLDExtraTrees holds roughly an 18\% lead over ExtraTrees$^*$. It implies that SPLDExtraTrees resistant mutations predictions are of higher precision and are composed of less false positives compared with ExtraTrees$^*$. In the case of interpolation, AP of SPLDExtraTrees is consistent with Rosetta (REF15) and is superior to molecular dynamics simulations (A99), which implies that training the proposed method on the tyrosine kinase-related data can achieve remarkable performance. Furthermore, we found that the additional physical and structural features could provide valuable information to improve AUC and ROC performance, the detail information see Figures S10-S13 in the supplementary material.

\section{Conclusion}
In this paper, we propose a robust machine learning method for predicting ligand binding affinity changes upon protein mutations for cancer target Abl kinase and identifying resistant mutations. Instead of feeding all samples into the training pool at once, the proposed method starts with learning both the easy and diverse samples and gradually takes more complex samples into consideration. By combining self-paced learning and the extremely random regression trees, the proposed method can quickly grasp easy and comprehensive knowledge and improve generalization performance. Furthermore, the diverse self-paced learning regularization terms can be used to incorporate the domain knowledge for the model training. This hypothesis is substantiated by our experiments. We conduct experiments on the TKI dataset under three scenarios to verify the effectiveness of the proposed method for predicting drug response to tyrosine kinase resistance. Our empirical results demonstrate that the proposed method outperforms molecular dynamics simulations with the Amber force fields and the traditional extremely random regression trees by a large margin across all scenarios. Although the Pearson correlation coefficient of the proposed method performs slightly worse than Rosetta REF15 scoring function in Scenario 1 (i.e., no tyrosine kinase in the training set), the proposed method could improve its ability to predict affinity changes in Abl mutants by learning a small amount of tyrosine kinase information. In particular, when the proposed method is trained on the most relevant data (i.e., TKI dataset), it achieves the best performance among all competing computational methods. 

The above results demonstrate that a good combination of domain knowledge and machine learning methods could greatly improve the prediction accuracy and generalization ability. Here, we mainly use protein family information in SPLDExtraTrees as a standard to select samples during the model training process. We also consider to use the amino acid (aa) change types (i.e., hydrophobic aa to polar aa, see Table S2 in the supplementary material for more information) as a standard in SPLDExtraTrees. The aa change types could also improve the predictive performance by a small margin in Scenario 1. However, it is much less obvious when compared with the protein family information. Interestingly, we find that for most of the aa change types we obtained good predictions, but failed to scale well only for the ``polar to hydrophobic'' type (see Figure S14 in the supplementary material for detail). This suggests that SPLDExtraTrees can still learn from aa change types, although classifying samples by the protein family seems to convey more useful information pertaining to the binding free energy changes.

We test the proposed methods on TKI dataset, which is an Abl kinase mutation dataset. We expect SPLDExtraTrees would effectively provide routine predictions of resistance-causing mutations for kinase. In addition, we mainly discussed single-point mutations where one aa was replaced with another aa in this work, but the prediction for more challenging site-mutation types (e.g., insertions, deletions, framework changes) and multi-point mutations need to be explored in the future work. We also expect our methods could be applied to these scenarios as solutions; however, appropriate datasets and incorporation of additional domain knowledge, through feature engineering or model training objectives, might be required.

\section{Supporting information Available}
The Supplementary Material is available on the ``Supplementary Material.docx'' file. \\
\textbf{Data S1:} Detailed information on the data set and numerical results for all calculations, provided as an Excel spreadsheet (XLSX).\\
\textbf{Data S2:} Input files used for the training and testing of the machine learning methods, provided as an Excel spreadsheet (XLSX).

\bibliographystyle{unsrt}  
\bibliography{references}

\begin{thebibliography}{10}

\bibitem{lovly2014molecular}
Christine~M Lovly and Alice~T Shaw.
\newblock Molecular pathways: resistance to kinase inhibitors and implications
  for therapeutic strategies.
\newblock {\em Clinical Cancer Research}, 20(9):2249--2256, 2014.

\bibitem{housman2014drug}
Genevieve Housman, Shannon Byler, Sarah Heerboth, Karolina Lapinska, Mckenna
  Longacre, Nicole Snyder, and Sibaji Sarkar.
\newblock Drug resistance in cancer: an overview.
\newblock {\em Cancers}, 6(3):1769--1792, 2014.

\bibitem{ward2020challenges}
Richard~A Ward, Stephen Fawell, Nicolas Floc’h, Vikki Flemington, Darren
  McKerrecher, and Paul~D Smith.
\newblock Challenges and opportunities in cancer drug resistance.
\newblock {\em Chemical Reviews}, 121(6):3297--3351, 2020.

\bibitem{aldeghi2019predicting}
Matteo Aldeghi, Vytautas Gapsys, and Bert~L de~Groot.
\newblock Predicting kinase inhibitor resistance: physics-based and data-driven
  approaches.
\newblock {\em ACS central science}, 5(8):1468--1474, 2019.

\bibitem{zehir2017mutational}
Ahmet Zehir, Ryma Benayed, Ronak~H Shah, Aijazuddin Syed, Sumit Middha,
  Hyunjae~R Kim, Preethi Srinivasan, Jianjiong Gao, Debyani Chakravarty, Sean~M
  Devlin, et~al.
\newblock Mutational landscape of metastatic cancer revealed from prospective
  clinical sequencing of 10,000 patients.
\newblock {\em Nature medicine}, 23(6):703--713, 2017.

\bibitem{fowler2018robust}
Philip~W Fowler, Kevin Cole, N~Claire Gordon, Angela~M Kearns, Martin~J
  Llewelyn, Tim~EA Peto, Derrick~W Crook, and A~Sarah Walker.
\newblock Robust prediction of resistance to trimethoprim in staphylococcus
  aureus.
\newblock {\em Cell chemical biology}, 25(3):339--349, 2018.

\bibitem{bhullar2018kinase}
Khushwant~S Bhullar, Naiara~Orrego Lagar{\'o}n, Eileen~M McGowan, Indu Parmar,
  Amitabh Jha, Basil~P Hubbard, and HP~Vasantha Rupasinghe.
\newblock Kinase-targeted cancer therapies: progress, challenges and future
  directions.
\newblock {\em Molecular cancer}, 17(1):1--20, 2018.

\bibitem{hauser2018predicting}
Kevin Hauser, Christopher Negron, Steven~K Albanese, Soumya Ray, Thomas
  Steinbrecher, Robert Abel, John~D Chodera, and Lingle Wang.
\newblock Predicting resistance of clinical abl mutations to targeted kinase
  inhibitors using alchemical free-energy calculations.
\newblock {\em Communications biology}, 1(1):1--14, 2018.

\bibitem{roskoski2021properties}
Robert Roskoski.
\newblock Properties of fda-approved small molecule protein kinase inhibitors:
  a 2021 update.
\newblock {\em Pharmacological research}, page 105463, 2021.

\bibitem{arora2005role}
Amit Arora and Eric~M Scholar.
\newblock Role of tyrosine kinase inhibitors in cancer therapy.
\newblock {\em Journal of Pharmacology and Experimental Therapeutics},
  315(3):971--979, 2005.

\bibitem{pottier2020tyrosine}
Charles Pottier, Margaux Fresnais, Marie Gilon, Guy J{\'e}rusalem, R{\'e}mi
  Longuesp{\'e}e, and Nor~Eddine Sounni.
\newblock Tyrosine kinase inhibitors in cancer: breakthrough and challenges of
  targeted therapy.
\newblock {\em Cancers}, 12(3):731, 2020.

\bibitem{weisberg2007second}
Ellen Weisberg, Paul~W Manley, Sandra~W Cowan-Jacob, Andreas Hochhaus, and
  James~D Griffin.
\newblock Second generation inhibitors of bcr-abl for the treatment of
  imatinib-resistant chronic myeloid leukaemia.
\newblock {\em Nature Reviews Cancer}, 7(5):345--356, 2007.

\bibitem{y2011recent}
X~Y~Lu, Q~Cai, and K~Ding.
\newblock Recent developments in the third generation inhibitors of bcr-abl for
  overriding t315i mutation.
\newblock {\em Current medicinal chemistry}, 18(14):2146--2157, 2011.

\bibitem{juchum2015fighting}
Michael Juchum, Marcel G{\"u}nther, and Stefan~A Laufer.
\newblock Fighting cancer drug resistance: Opportunities and challenges for
  mutation-specific egfr inhibitors.
\newblock {\em Drug Resistance Updates}, 20:12--28, 2015.

\bibitem{neel2017resistance}
Dana~S Neel and Trever~G Bivona.
\newblock Resistance is futile: overcoming resistance to targeted therapies in
  lung adenocarcinoma.
\newblock {\em NPJ precision oncology}, 1(1):1--6, 2017.

\bibitem{patel2017mechanisms}
Ami~B Patel, Thomas O’Hare, and Michael~W Deininger.
\newblock Mechanisms of resistance to abl kinase inhibition in chronic myeloid
  leukemia and the development of next generation abl kinase inhibitors.
\newblock {\em Hematology/Oncology Clinics}, 31(4):589--612, 2017.

\bibitem{gapsys2015pmx}
Vytautas Gapsys, Servaas Michielssens, Daniel Seeliger, and Bert~L de~Groot.
\newblock pmx: Automated protein structure and topology generation for
  alchemical perturbations, 2015.

\bibitem{wang2015accurate}
Lingle Wang, Yujie Wu, Yuqing Deng, Byungchan Kim, Levi Pierce, Goran Krilov,
  Dmitry Lupyan, Shaughnessy Robinson, Markus~K Dahlgren, Jeremy Greenwood,
  et~al.
\newblock Accurate and reliable prediction of relative ligand binding potency
  in prospective drug discovery by way of a modern free-energy calculation
  protocol and force field.
\newblock {\em Journal of the American Chemical Society}, 137(7):2695--2703,
  2015.

\bibitem{steinbrecher2015accurate}
Thomas~B Steinbrecher, Markus Dahlgren, Daniel Cappel, Teng Lin, Lingle Wang,
  Goran Krilov, Robert Abel, Richard Friesner, and Woody Sherman.
\newblock Accurate binding free energy predictions in fragment optimization.
\newblock {\em Journal of chemical information and modeling},
  55(11):2411--2420, 2015.

\bibitem{aldeghi2018accurate}
Matteo Aldeghi, Vytautas Gapsys, and Bert~L de~Groot.
\newblock Accurate estimation of ligand binding affinity changes upon protein
  mutation.
\newblock {\em ACS central science}, 4(12):1708--1718, 2018.

\bibitem{alford2017rosetta}
Rebecca~F Alford, Andrew Leaver-Fay, Jeliazko~R Jeliazkov, Matthew~J O’Meara,
  Frank~P DiMaio, Hahnbeom Park, Maxim~V Shapovalov, P~Douglas Renfrew,
  Vikram~K Mulligan, Kalli Kappel, et~al.
\newblock The rosetta all-atom energy function for macromolecular modeling and
  design.
\newblock {\em Journal of chemical theory and computation}, 13(6):3031--3048,
  2017.

\bibitem{barlow2018flex}
Shane Barlow, Kyle~A, Samuel Thompson, Pooja Suresh, James~E Lucas, Markus
  Heinonen, and Tanja Kortemme.
\newblock Flex ddg: Rosetta ensemble-based estimation of changes in
  protein--protein binding affinity upon mutation.
\newblock {\em The Journal of Physical Chemistry B}, 122(21):5389--5399, 2018.

\bibitem{shu2019meta}
Jun Shu, Qi~Xie, Lixuan Yi, Qian Zhao, Sanping Zhou, Zongben Xu, and Deyu Meng.
\newblock Meta-weight-net: Learning an explicit mapping for sample weighting.
\newblock {\em arXiv preprint arXiv:1902.07379}, 2019.

\bibitem{yang2020smspl}
Ziyi Yang, Naiqi Wu, Yong Liang, Hui Zhang, and Yanqiong Ren.
\newblock Smspl: Robust multimodal approach to integrative analysis of
  multiomics data.
\newblock {\em IEEE Transactions on Cybernetics}, 2020.

\bibitem{kumar2010self}
M~Pawan Kumar, Benjamin Packer, and Daphne Koller.
\newblock Self-paced learning for latent variable models.
\newblock In {\em NIPS}, volume~1, page~2, 2010.

\bibitem{jiang2014easy}
Lu~Jiang, Deyu Meng, Teruko Mitamura, and Alexander~G Hauptmann.
\newblock Easy samples first: Self-paced reranking for zero-example multimedia
  search.
\newblock In {\em Proceedings of the 22nd ACM international conference on
  Multimedia}, pages 547--556, 2014.

\bibitem{jiang2014self}
Lu~Jiang, Deyu Meng, Shoou-I Yu, Zhenzhong Lan, Shiguang Shan, and Alexander
  Hauptmann.
\newblock Self-paced learning with diversity.
\newblock In {\em Advances in Neural Information Processing Systems}, pages
  2078--2086, 2014.

\bibitem{yang2019multi}
Zi-Yi Yang, Xiao-Ying Liu, Jun Shu, Hui Zhang, Yan-Qiong Ren, Zong-Ben Xu, and
  Yong Liang.
\newblock Multi-view based integrative analysis of gene expression data for
  identifying biomarkers.
\newblock {\em Scientific reports}, 9(1):1--15, 2019.

\bibitem{pires2015platinum}
Douglas~EV Pires, Tom~L Blundell, and David~B Ascher.
\newblock Platinum: a database of experimentally measured effects of mutations
  on structurally defined protein--ligand complexes.
\newblock {\em Nucleic acids research}, 43(D1):D387--D391, 2015.

\bibitem{schymkowitz2005foldx}
Joost Schymkowitz, Jesper Borg, Francois Stricher, Robby Nys, Frederic
  Rousseau, and Luis Serrano.
\newblock The foldx web server: an online force field.
\newblock {\em Nucleic acids research}, 33(suppl\_2):W382--W388, 2005.

\bibitem{salentin2015plip}
Sebastian Salentin, Sven Schreiber, V~Joachim Haupt, Melissa~F Adasme, and
  Michael Schroeder.
\newblock Plip: fully automated protein--ligand interaction profiler.
\newblock {\em Nucleic acids research}, 43(W1):W443--W447, 2015.

\bibitem{trott2010autodock}
Oleg Trott and Arthur~J Olson.
\newblock Autodock vina: improving the speed and accuracy of docking with a new
  scoring function, efficient optimization, and multithreading.
\newblock {\em Journal of computational chemistry}, 31(2):455--461, 2010.

\bibitem{wang2017improving}
Cheng Wang and Yingkai Zhang.
\newblock Improving scoring-docking-screening powers of protein--ligand scoring
  functions using random forest.
\newblock {\em Journal of computational chemistry}, 38(3):169--177, 2017.

\bibitem{liang2013sparse}
Yong Liang, Cheng Liu, Xin-Ze Luan, Kwong-Sak Leung, Tak-Ming Chan, Zong-Ben
  Xu, and Hai Zhang.
\newblock Sparse logistic regression with a l 1/2 penalty for gene selection in
  cancer classification.
\newblock {\em BMC bioinformatics}, 14(1):1--12, 2013.

\bibitem{dudoit2002comparison}
Sandrine Dudoit, Jane Fridlyand, and Terence~P Speed.
\newblock Comparison of discrimination methods for the classification of tumors
  using gene expression data.
\newblock {\em Journal of the American statistical association},
  97(457):77--87, 2002.

\bibitem{li2004comparative}
Tao Li, Chengliang Zhang, and Mitsunori Ogihara.
\newblock A comparative study of feature selection and multiclass
  classification methods for tissue classification based on gene expression.
\newblock {\em Bioinformatics}, 20(15):2429--2437, 2004.

\bibitem{lee2005extensive}
Jae~Won Lee, Jung~Bok Lee, Mira Park, and Seuck~Heun Song.
\newblock An extensive comparison of recent classification tools applied to
  microarray data.
\newblock {\em Computational Statistics \& Data Analysis}, 48(4):869--885,
  2005.

\bibitem{ding2005minimum}
Chris Ding and Hanchuan Peng.
\newblock Minimum redundancy feature selection from microarray gene expression
  data.
\newblock {\em Journal of bioinformatics and computational biology},
  3(02):185--205, 2005.

\bibitem{monari2000withdrawing}
Ga{\'e}tan Monari and G{\'e}rard Dreyfus.
\newblock Withdrawing an example from the training set: An analytic estimation
  of its effect on a non-linear parameterised model.
\newblock {\em Neurocomputing}, 35(1-4):195--201, 2000.

\bibitem{rivals2003mlps}
Isabelle Rivals and L{\'e}on Personnaz.
\newblock Mlps (mono layer polynomials and multi layer perceptrons) for
  nonlinear modeling.
\newblock {\em The Journal of Machine Learning Research}, 3:1383--1398, 2003.

\bibitem{golub1999molecular}
Todd~R Golub, Donna~K Slonim, Pablo Tamayo, Christine Huard, Michelle
  Gaasenbeek, Jill~P Mesirov, Hilary Coller, Mignon~L Loh, James~R Downing,
  Mark~A Caligiuri, et~al.
\newblock Molecular classification of cancer: class discovery and class
  prediction by gene expression monitoring.
\newblock {\em science}, 286(5439):531--537, 1999.

\bibitem{yang2018robust}
Zi-Yi Yang, Yong Liang, Hui Zhang, Hua Chai, Bowen Zhang, and Cheng Peng.
\newblock Robust sparse logistic regression with the $l_q$ ($0<q<1$)
  regularization for feature selection using gene expression data.
\newblock {\em IEEE Access}, 6:68586--68595, 2018.

\bibitem{bengio2009curriculum}
Yoshua Bengio, J{\'e}r{\^o}me Louradour, Ronan Collobert, and Jason Weston.
\newblock Curriculum learning.
\newblock In {\em Proceedings of the 26th annual international conference on
  machine learning}, pages 41--48, 2009.

\bibitem{das2015diversity}
Sayoni Das, Natalie~L Dawson, and Christine~A Orengo.
\newblock Diversity in protein domain superfamilies.
\newblock {\em Current opinion in genetics \& development}, 35:40--49, 2015.

\bibitem{yuan2006model}
Ming Yuan and Yi~Lin.
\newblock Model selection and estimation in regression with grouped variables.
\newblock {\em Journal of the Royal Statistical Society: Series B (Statistical
  Methodology)}, 68(1):49--67, 2006.

\bibitem{wang2004development}
Junmei Wang, Romain~M Wolf, James~W Caldwell, Peter~A Kollman, and David~A
  Case.
\newblock Development and testing of a general amber force field.
\newblock {\em Journal of computational chemistry}, 25(9):1157--1174, 2004.

\bibitem{hornak2006comparison}
Viktor Hornak, Robert Abel, Asim Okur, Bentley Strockbine, Adrian Roitberg, and
  Carlos Simmerling.
\newblock Comparison of multiple amber force fields and development of improved
  protein backbone parameters.
\newblock {\em Proteins: Structure, Function, and Bioinformatics},
  65(3):712--725, 2006.

\bibitem{best2009optimized}
Robert~B Best and Gerhard Hummer.
\newblock Optimized molecular dynamics force fields applied to the helix- coil
  transition of polypeptides.
\newblock {\em The journal of physical chemistry B}, 113(26):9004--9015, 2009.

\bibitem{lindorff2010improved}
Kresten Lindorff-Larsen, Stefano Piana, Kim Palmo, Paul Maragakis, John~L
  Klepeis, Ron~O Dror, and David~E Shaw.
\newblock Improved side-chain torsion potentials for the amber ff99sb protein
  force field.
\newblock {\em Proteins: Structure, Function, and Bioinformatics},
  78(8):1950--1958, 2010.

\end{thebibliography}

\end{document}